\documentclass[aps,prl,twocolumn, longbibliography,footinbib,superscriptaddress, preprintnumbers]{revtex4-1}

\usepackage{hyperref}
\usepackage{amsmath}
\usepackage{amssymb}
\usepackage{amsthm}
\usepackage{graphicx}
\usepackage{bbm}
\usepackage{CJK}

\renewcommand{\thanks}[1]{\footnote{#1}}

\makeatletter
\def\@hangfrom@section#1#2#3{\@hangfrom{#1#2}#3}
\def\@hangfroms@section#1#2{#1#2}
\makeatother

\newcommand{\bea}{\begin{eqnarray}}
\newcommand{\eea}{\end{eqnarray}}
\newcommand{\be}{\begin{eqnarray}}
\newcommand{\ee}{\end{eqnarray}}

\def\ie{\begin{equation}\begin{aligned}}
\def\fe{\end{aligned}\end{equation}}

\def\ie{\begin{equation}\begin{aligned}}
\def\fe{\end{aligned}\end{equation}}

\def\cI{{\cal I}}

\begin{document}
\begin{CJK*}{UTF8}{gbsn}

\preprint{
QMUL-PH-26-17
}

\title{Bootstrapping Giant Graviton Correlators}

\author{Song He (何颂)}
\email{songhe@itp.ac.cn}
\affiliation{New Cornerstone Laboratory, Institute of Theoretical Physics, Chinese Academy of Sciences, Beijing 100190, China}
\affiliation{School of Fundamental Physics and Mathematical Sciences, Hangzhou Institute for Advanced Study, UCAS, Hangzhou 310024, China} 
\author{Canxin Shi (施灿欣)}
\email{shicanxin@itp.ac.cn}
\affiliation{New Cornerstone Laboratory, Institute of Theoretical Physics, Chinese Academy of Sciences, Beijing 100190, China}
\author{Yichao Tang (唐一朝)}
\email{tangyichao@itp.ac.cn}
\affiliation{New Cornerstone Laboratory, Institute of Theoretical Physics, Chinese Academy of Sciences, Beijing 100190, China}
\affiliation{School of Physical Sciences, University of Chinese Academy of Sciences, Beijing 100049, China}
\author{Congkao Wen (温从烤)}
\email{c.wen@qmul.ac.uk}
\affiliation{Centre for Theoretical Physics and Astronomy, Department of Physics and Astronomy, Queen Mary University of London, London, E1 4NS, UK}

\begin{abstract}
We develop bootstrap methods for mixed heavy-light four-point correlators $\langle GGOO\rangle$ in $\mathcal N=4$ super-Yang--Mills theory at large $N$, where $O\equiv {\cal O}_2$ is the chiral primary operator in the stress-tensor multiplet and $G$ are (dual) giant graviton operators with dimension of order $N$, including the maximal determinant case. The loop integrand is expanded in a basis of labelled $f$-graphs -- necessarily including non-planar topologies due to the dimension-$N$ nature of the giant gravitons -- and the coefficients are fixed by various bootstrap conditions including double-triangle and triangle rules in the cusp and OPE limits, integrated correlators from supersymmetric localization, and a ten-dimensional hidden symmetry, the latter also allowing extension to correlators involving generic chiral primaries $\mathcal{O}_k$. Together, these inputs uniquely determine the correlator through three loops, passing further non-trivial consistency checks. For the maximal determinant operator, we reproduce the known results through two loops and obtain the full three-loop correction.
\end{abstract}

\maketitle
\end{CJK*}
\section{Introduction}

Giant gravitons are a distinguished class of half-BPS operators in $\mathcal{N}=4$
super-Yang--Mills theory (SYM) whose conformal dimensions are of order $N$, placing
them far outside the familiar single-trace sector.
In the AdS/CFT correspondence they are dual to D3-branes wrapping cycles in
$AdS_5\times S^5$~\cite{McGreevy:2000cw,Hashimoto:2000zp,Corley:2001zk}; their correlation functions probe a regime where
large-$N$ counting and perturbative dynamics interplay in ways not accessible to
standard planar methods.
Extending bootstrap techniques to these heavy operators presents qualitatively new
challenges: the reduced symmetry enlarges the space of admissible integrands, 
non-planar topologies contribute even at leading large $N$, and the relevant OPE data involving giant gravitons is less known.
Recent developments -- including powerful graphical bootstrap methods, all-order
results for integrated correlators from supersymmetric localization, and a hidden
ten-dimensional symmetry -- have nevertheless brought such correlators within reach
of systematic perturbative analysis.

We study mixed heavy--light four-point correlators $\langle GGOO
\rangle$, with two (dual) giant graviton operators $G$ and two chiral primary operators of
fixed dimensions.
The giant gravitons are conveniently parameterised by $\alpha$: for $0 < \alpha \leq 1$ they are the sphere giant gravitons, with $\alpha = 1$ the full determinant operator, while $\alpha <0$ gives dual giant gravitons.
For the light operators we focus on the dimension-two chiral primary $\mathcal{O}_2$ in the stress-tensor multiplet; a recently proposed ten-dimensional hidden symmetry~\cite{Chen:2025yxg, Wu:2025ott} then allows the results to be promoted to chiral primaries of generic dimension. 

To implement the bootstrap, we expand the loop integrand in a basis of labelled
$f$-graphs~\cite{Eden:2011we}, crucially including non-planar topologies.
The coefficients in this ansatz are then fixed by several complementary inputs:
double-triangle relations from light-like cusp limits~\cite{He:2024cej}, triangle rules in the $GG$,
$OO$, and $GO$ OPE channels, integrated correlator constraints~\cite{Brown:2024tru, Brown:2025huy, Brown:2026dhy} from supersymmetric
localization, and a conjectured extension of the ten-dimensional hidden symmetry to
generic giant graviton operators.
We further exploit the fact that OPE coefficients in $\mathcal{N}=4$ SYM admit
representations in terms of harmonic sums~\cite{Eden:2000bk}, providing additional
constraints.

Together these (over-constrained) conditions uniquely determine the correlator through three loops.
For $\alpha=1$ we reproduce the known answer through two loops~\cite{Jiang:2019xdz} and provide the complete 
three-loop result. 
We extract new OPE data throughout: in particular, the OPE coefficients $d_S$ of
two giant gravitons and a twist-two operator of spin $S$ generically grow as
$(\log S)^\ell$ at $\ell$ loops, but remarkably reduce to $\log S$ for the maximal
determinant operator with $\alpha=1$, confirming and extending an observation made at two
loops~\cite{Jiang:2019xdz}.
The results exhibit further characteristic structure: enhanced permutation
symmetry in the planar $f$-graphs at $\alpha=-1$, and in the very-heavy dual giant
graviton limit $\alpha\to-\infty$, the correlator is dominated by a special class of
$f$-graphs, in agreement with the expected semi-classical description. 

\subsection{Bootstrap setup}

The sphere and AdS giant gravitons are extracted from the generating functions as 
{\allowdisplaybreaks\begin{gather}
   G_{\mathrm{S}; \alpha}(x, y):=\oint \frac{d z}{2 \pi \mathrm{i}} \frac{1}{z^{\alpha N+1}} \operatorname{det}(\mathbbm{1}+z\, y \cdot \Phi(x)) \, , \\
   G_{\mathrm{AdS}; \beta}(x, y):=\oint \frac{d z}{2 \pi \mathrm{i}} \frac{1}{z^{\beta N+1}} \frac{1}{\operatorname{det}(\mathbbm{1}-z\, y \cdot \Phi(x))} \, ,
\end{gather}}
where $\Phi^I(x)$ with $I=1,\ldots,6$ denotes the six scalars of $\mathcal{N}=4$ SYM, and $y_I$ is a null SO(6) polarisation vector. The parameters satisfy $0 < \alpha \leq 1$ and $\beta > 0$. Inspired by the symmetry $\alpha \leftrightarrow -\beta$ of the integrated correlators observed in~\cite{Brown:2025huy}, we use the unified notation 
\begin{equation}
G_{\alpha} := \left\{ \begin{array}{ll}
   G_{\mathrm{S}; \alpha} & ~ \text{for }~ 0 < \alpha \leq 1\, , \\
   G_{\mathrm{AdS}; -\alpha} & ~ \text{for } ~ \alpha < 0\, , 
\end{array}\right.
\end{equation}
for the two types of giant gravitons. The four-point correlator we will consider is defined as
\begin{equation}
\!\!\! \! \mathcal{T}_{\alpha} :=\! \frac{\langle G_{\alpha}(x_1,y_1) G_{\alpha}(x_2,y_2) \mathcal{O}_2(x_3,y_3) \mathcal{O}_2(x_4,y_4) \rangle}{\langle G_{\alpha}(x_1,y_1) G_{\alpha}(x_2,y_2) \rangle \langle \mathcal{O}_2(x_3,y_3) \mathcal{O}_2(x_4,y_4) \rangle} \, ,
\end{equation}
where $\mathcal{O}_2(x,y) = \operatorname{Tr}[ \left(y \cdot \Phi\left(x\right)\right)^{2} ]$ is the chiral primary operator. In the weak-coupling expansion, they can be expressed as~\cite{Eden:2000bk, Nirschl:2004pa}
\begin{equation}
\mathcal{T}_{\alpha} = \mathcal{T}^{\rm free}_{\alpha} +\frac{4 |\alpha|\,  R_{1234}}{ N d_{12}^{2} d_{34}^2 x_{13}^2 x_{24}^2} \sum_{\ell=1}^\infty a^\ell \, \mathcal{T}^{(\ell)}_{\alpha} \, ,
\end{equation}
where we have defined 
\begin{equation}
a := \frac{g_{\rm YM}^2N}{4\pi^2},
\quad x_{ij}^2 := (x_i - x_j)^2, 
\quad d_{ij} := \frac{2\, y_i \!\cdot\! y_j }{x_{ij}^2} \, , 
\end{equation} 
and the prefactor $R_{1234}$ is fixed by the superconformal symmetry.  We will now implement the bootstrap construction of the correlator.    

\paragraph{Ansatz}  We begin by introducing the basis for the loop integrands of the correlator. The $\ell$ loop contribution can be expressed as
\begin{equation}
   \label{eq:reducedCorrelator}
\mathcal{T}^{(\ell)}_{\alpha} = \frac{\xi_4}{\ell!} \int \frac{\mathrm{d}^4 x_5 \cdots \mathrm{~d}^4 x_{4+\ell}}{\left(-4 \pi^2\right)^{\ell}} F_{4+\ell}\, ,
\end{equation}
where $\xi_4 = x_{13}^4 x_{24}^4 x_{12}^2 x_{23}^2 x_{34}^2 x_{41}^2$. 
The integrand $F_{4+\ell}$ can be constructed through the chiral Lagrangian-insertion procedure, as has been done for $\langle \mathcal{O}_2 \mathcal{O}_2 \mathcal{O}_2 \mathcal{O}_2 \rangle$~\cite{Eden:2011we,Eden:2012tu}. In this way, $F_{4+\ell}$ can be expressed in terms of $f$-graphs, however, unlike the case of $\langle \mathcal{O}_2 \mathcal{O}_2 \mathcal{O}_2 \mathcal{O}_2 \rangle$, due to the insertion of two giant gravitons,  $F_{4+\ell}$ has reduced permutation symmetry $S_2 \times S_{2+\ell}$; we will denote such more general $f$-graphs as \emph{labelled} $f$-graphs~\cite{Jiang:2023uut}. In conclusion, we have
\begin{gather}
F_{4+\ell}=\sum_{n} c^{(\ell)}_n f^{(\ell)}_n \, , \label{eq:F-graph} \\
f^{(\ell)}_n = \frac{p^{(\ell)}_n(x_i)}{\prod_{1\leq i < j \leq 4+\ell} x_{ij}^2} +P_{12;34\cdots (4+\ell)} \, , \label{eq:p-numerator}
\end{gather}
where $P_{12;34\cdots (4+\ell)}$ denotes permutations of $\{1,2\}$ and $\{3,4,\cdots, 4+\ell\}$, and the numerator $p^{(\ell)}_n(x_i)$ is a monomial in $x_{ij}^2$ of degree $(\ell{-}1)$ at each point $x_i$. For convenience, we normalize it so that after summing over the permutations, the coefficient of each term is 1. 
$f^{(\ell)}_n$ can be represented by $(4\!+\!\ell)$-point graphs: vertices denote the points $x_i$, with $x_1$ and $x_2$ distinguished to reflect the permutation structure; solid and dashed lines denote factors of $1/x_{ij}^2$ and $x_{ij}^2$, respectively. One- and two-loop examples are shown in Figure~\ref{fig:fgraph-1loop}; see the Appendix~\ref{app:f-graphs} for the explicit three-loop graph basis and related technical details.
\nocite{Brown:2024yvt,Brown:2025cbz,Caetano:2023zwe,Coronado:2025xwk,Dorigoni:2021guq,Giombi:2020enj,HyperlogProcedures}
\begin{figure}[t]
\centering
\begin{tabular}{c}
\includegraphics[width=0.15\textwidth]{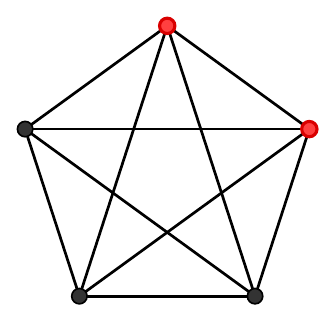} \\
$f_1^{(1)}$
\end{tabular}

\begin{tabular}{cc}
\includegraphics[width=0.15\textwidth]{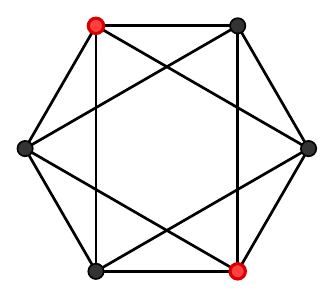} &\hspace{8pt}
\includegraphics[width=0.15\textwidth]{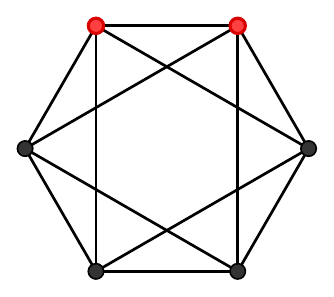} \\
$f_1^{(2)}$ &\hspace{8pt} $f_2^{(2)}$
\end{tabular}
\caption{Labelled $f$-graphs at one and two loops, where each solid line represents a propagator $1/x_{ij}^2$. The highlighted vertices indicate the distinguished points $x_1$ and $x_2$.}
\label{fig:fgraph-1loop}
\end{figure}
Importantly, the topology of the $f$-graphs (defined by the subgraph containing only solid lines) is not necessarily planar: the intricate  structure and dimension-$N$ nature of giant graviton operators allow non-planar graphs to contribute even in the large-$N$ limit. The coefficients $c_n^{(\ell)}$ are fixed by imposing several constraints, each of which yields linear relations among them. We now briefly summarize these constraints.

\paragraph{Double-triangle rule} 
The double-triangle rule originates from the behavior of the correlator $\langle \mathcal{O}_2 \mathcal{O}_2 \mathcal{O}_2 \mathcal{O}_2 \rangle$ in the ``cusp'' limit, where three points become consecutively light-like separated, i.e. $x_{ij}^2, x_{jk}^2 \to 0$~\cite{He:2024cej, Bourjaily:2025iad}. 
At the level of the integrand, the double-triangle rule takes the form 
\begin{equation}
\label{eq:doubleTriangle}
\lim_{\substack{x_{ij}^2, x_{jk}^2 \rightarrow 0 \\ x_{n} \rightarrow x_j}} \frac{x_{ij}^2 x_{jk}^2 x_{i, n}^2 x_{j, n}^2 x_{k, n}^2}{x_{ik}^2} F_{n} -2\, x_{ij}^2 x_{jk}^2 F_{n-1} =0 \, . 
\end{equation}
It is straightforward to generalize this rule to giant graviton correlators. There are different kinds of cusp singularities depending on which operators participate in the cusp. We consider two kinds of double-triangle rules: 
\begin{itemize}
   \item $GOO$ cusp, with $x_{23}^2, x_{34}^2 \to 0$\, ;
   \item $LOO$ cusp, with $x_{53}^2, x_{34}^2 \to 0$\, .
\end{itemize}
Here $L$ denotes a Lagrangian-insertion point; in the example above this is the inserted point $x_5$.

\paragraph{Triangle rule} 
The triangle rule is a graphical representation of the constraints imposed by the OPE limit~\cite{Bourjaily:2016evz}. In the case of the limit $x_i \to x_j$, the triangle rule for the integrand can be expressed as
\begin{equation}
\lim _{x_j, x_{n} \rightarrow x_i} x_{ij}^2 x_{i, n}^2 x_{j, n}^2 F_{n} - 2 \gamma_{\mathcal{O}}\, x_{ij}^2 F_{n-1} = 0\, , 
\end{equation}
where $\gamma_{\mathcal{O}}$ is the one-loop anomalous dimension of the exchanged operator that dominates the OPE limit. Because we have two kinds of operators, there are three different OPE channels:
\begin{itemize}
   \item $GG$ channel, with $x_{2} \to x_1$: the dominant contribution is from the Konishi operator $\mathcal{K}$, with $\gamma_{\mathcal{K}} = 3$~\cite{Fiamberti:2007rj};
   \item $OO$ channel, with $x_{4} \to x_3$: the dominant contribution is also from the Konishi operator;
   \item $GO$ channel, with $x_{4} \to x_1$: the dominant contribution is from an unknown operator $G'$, whose anomalous dimension can be fixed using the triangle rule at two loops, yielding $\gamma_{G'} = 2-\alpha$~\footnote{As we will show, this condition is not needed to determine the correlators up to two loops; it first becomes useful at three loops. We also assume that there is a single unprotected operator (or several operators with $\langle\gamma_{G'}\rangle=2-\alpha$) that dominates the OPE limit.}.  
\end{itemize}

\paragraph{Integrated-correlator constraints} 
As shown in~\cite{Wen:2022oky,Brown:2023zbr}, the integrated correlators~\cite{Binder:2019jwn, Chester:2020dja}, when applied to the perturbative expansion of the correlator, can be expressed as periods of $f$-graphs:
\begin{equation} \label{eq:Int}
   \cI_{\alpha}(a) = -\sum_{\ell=1}^{\infty} \frac{4 |\alpha|}{\ell!} \left(-\frac{a}{4}\right)^{\ell} \sum_{n} c_n^{(\ell)} \, \mathcal{P} \big(f^{(\ell)}_n \big) \, ,
\end{equation}
where the periods of $f$-graphs are defined as
\begin{equation}
\mathcal{P} \big(f^{(\ell)}_n \big) = \frac{1}{(\pi^2)^{\ell+1}} \int \frac{d^4 x_1 \ldots d^4 x_{\ell+4}}{{\rm vol\, (SO(1, 5)}) }  f^{(\ell)}_n(x_i) \, . 
\end{equation}
Their explicit values through three loops are given in Appendix~\ref{app:periods}. For convenience, we quote the giant graviton integrated correlators through three loops, as obtained from the supersymmetric localization~\cite{Brown:2025huy},
\begin{align} \label{eq:integrated}
    \cI_{\alpha}(a) =&\, 6 \alpha\, \zeta (3)\, a + \frac{15}{2} \alpha  (\alpha - 4)\,  \zeta (5)\, a^2 \\
    & +\frac{35}{8} \alpha  \left(2 \alpha^2 - 15 \alpha +30\right) \zeta (7)\, a^3  +\dots \, . \nonumber
\end{align}
Given the periods of the $f$-graph integrals, we can compute the integrated correlators from our ansatz, and matching with the results from supersymmetric localization then imposes another set of constraints. 

\paragraph{Extraction of twist-two OPE data} 
A further source of constraints comes from extracting the OPE data of the twist-two trajectory $\mathcal{O}_S$ with arbitrary spin $S$ in the $\mathbf{20'}$ R-symmetry channel. In the double OPE limit $x_2 \to x_1, x_4 \to x_3$, its contribution takes the schematic form~\cite{Dolan:2000ut,Eden:2012rr}     
\begin{equation}
\label{eq:twist2OPEinput}
\left.\mathcal{T}_{\alpha}\right|_{\text{twist-2}}
 = \sum_{S} d_S \, c_S \, g_{\Delta_S,S}(z, \bar{z})+\cdots \, ,
\end{equation}
where $g_{\Delta_S,S}(z, \bar{z})$ is the conformal block with $\Delta_S=2+S+\gamma_S$, and $d_S$, $c_S$ denote the OPE coefficients of $\mathcal{O}_S$ with two giant gravitons and two chiral primary operators, respectively. Starting from the correlator ansatz with arbitrary $f$-graph coefficients, one can evaluate the corresponding conformal integrals and match the OPE expansion order by order. This allows us to extract the anomalous dimensions $\gamma_S$ and structure constants $d_S$ as functions of those coefficients, while also imposing nontrivial constraints on the $f$-graph coefficients.

\paragraph{Harmonic-sum representability}
Once the twist-two OPE data have been extracted for sufficiently many spins, one can further constrain the correlator by demanding that they admit a closed-form representation in terms of nested harmonic sums (denoted as $S_{\vec{a}}(S)$),
\begin{equation}
\label{eq:harmonicSumInput}
\gamma_S=\sum_{\vec{a}} c_{\vec{a}}\, S_{\vec{a}}(S)\, , \qquad
d_S=\sum_{\vec{a}} b_{\vec{a}}\, S_{\vec{a}}(S)\, ,
\end{equation}
with the sums taken over a basis of harmonic sums of the appropriate transcendental weight~\cite{Eden:2000bk}. In practice, we use the package \texttt{HarmonicSums}~\cite{Ablinger:2014rba} to construct this basis, and then fix the coefficients by matching to the OPE data extracted at finite spin~\cite{Eden:2012rr,Jiang:2019xdz,Jiang:2023uut}. Requiring that the data fit a single harmonic-sum expression for arbitrary spin then gives additional constraints on the remaining $f$-graph coefficients. Therefore, harmonic-sum representability is an extra bootstrap input, rather than merely a compact way of presenting the final answer.

\paragraph{Hidden symmetry}
It was recently proposed that the heavy-light correlators with maximal-determinant operators enjoy a ten-dimensional hidden symmetry~\cite{Chen:2025yxg, Wu:2025ott, Chen:2026ium} by extending what was known for four-point correlators of single-trace operators~\cite{Caron-Huot:2021usw, Caron-Huot:2023wdh}. This was achieved by viewing the heavy-light correlators as defect systems, a perspective that naturally extends to more generic (dual) giant gravitons. We conjecture the hidden symmetry applies to heavy-light correlators with more general giant graviton operators considered here, supported by an explicit calculation at two loops~\cite{Coronado:2026}. 
We further impose the constraint, motivated by~\cite{Brown:2023zbr} for four-point correlators of single-trace operators that after summing over \(R\)-symmetry channels at \(\ell\) loops, all transcendental constants cancel except \(\zeta(2\ell{+}1)\); we assume this pattern extends to giant graviton correlators. 

\subsection{Giant graviton correlators up to two loops} 
At the tree level, the correlator reads 
\begin{multline} \label{eq:GGOOtree}
\mathcal{T}_{\alpha}^{\text{tree}}
= \frac{2 |\alpha|}{N d_{12}^2 d_{34}^2} \big( (|\alpha| N {-} 1) d_{13} d_{14} d_{23} d_{24} \\
+ d_{12} d_{14} d_{23} d_{34} + d_{12} d_{13} d_{24} d_{34}
\big) + O \Big(\frac{1}{N^2}\Big)\, ,
\end{multline}
which is obtained from explicit Wick contractions, and we note the result contains an $O(N^0)$ contribution.

At one loop, there is a single $f$-graph, namely the one-loop box integral; see also Figure~\ref{fig:fgraph-1loop}. We therefore only need to determine its overall coefficient, which is fixed by the integrated-correlator result \eqref{eq:integrated}. Using the period of the one-loop $f$-graph given in Appendix~\ref{app:periods}, we find
\begin{equation}
   \label{eq:1-loop-cor}
   F_{5} = f^{(1)}(x_i) \, .
\end{equation}
At two loops, there are two independent $f$-graphs, whose definitions are given in \eqref{eq:2-loop}.
The integrated-correlator constraint alone is not enough to determine them. Combining it with an OPE limit, for example the $OO$-channel OPE~\footnote{Since the $OO$-channel OPE only involves the anomalous dimension of the Konishi operator, which does not receive \(1/N\) corrections up to three loops~\cite{Fiamberti:2007rj}, this analysis also applies at finite \(N\), provided one uses the corresponding finite-\(N\) integrated correlators, as in~\cite{Brown:2026dhy}.}, we find 
\begin{equation} 
   \label{eq:2-loop-cor}
   F_6 = f^{(2)}_{2}(x_i) - \alpha f^{(2)}_{1}(x_i) \, . 
\end{equation}
We have also checked that this result is consistent with the double-triangle rules in the $GOO$ and $LOO$ cusp limits; and for $\alpha=1$ it reproduces the known maximal-determinant result of~\cite{Jiang:2019xdz, Jiang:2023uut}. A few further comments are in order. First, when $\alpha=-1$ (i.e. for a dual giant graviton operator of conformal dimension $N$), the $S_2 \times S_4$ permutation symmetry of the two-loop correlator is enhanced to $S_{6}$. Second, in the limit $\alpha \rightarrow - \infty$, the term $f^{(2)}_{1}(x_i)$ dominates, in agreement with expectations for the very-heavy dual giant graviton limit~\cite{Brown:2025huy}. Third, the one- and two-loop results determine the anomalous dimension in the $GO$-channel OPE to be $\gamma_{G'} = 2-\alpha$, as noted above. This result will be used in the three-loop bootstrap. Furthermore, assuming the ten-dimensional hidden symmetry~\cite{Chen:2025yxg, Wu:2025ott, Chen:2026ium}, our result provides the generating function for the correlator $\langle G\,G\, \mathcal{O}_{p} \mathcal{O}_{p+2k} \rangle$ for generic $p$ and $k$~\footnote{The result is in agreement with an independent computation 
of~\cite{Coronado:2026}. As commented earlier, this confirms the 10-dimensional hidden symmetry for generic giant gravitons at two loops.}.

\subsection{Giant graviton correlators  at three loops} 

We now consider the correlators at three loops. There are 15 labelled $f$-graphs at this order, which we denote by $f^{(3)}_{i,j}$ following the organization of Appendix~\ref{app:f-graphs}, where their explicit definitions are given. One Gram-determinant relation among the 15 integrals leaves 
14 independent ones; and we choose a basis in which the coefficient of $f^{(3)}_{1,2}$ is set to one. 

The number of independent coefficients fixed by each input is summarized in Table~\ref{tab:3loopConstraintCount}. We find two minimal sets of constraints that determine all but one of the independent coefficients. In both sets, the $GO$-channel OPE and integrated-correlator constraints are indispensable. By contrast, the cusp constraints $GOO{+}LOO$ and the twist-two constraints $\mathrm{Tw2}{+}\mathrm{HS}$ are interchangeable: either pair, combined with $GO$ and $\mathrm{Int}$, leaves only a single parameter unfixed before imposing hidden symmetry. The agreement between these two routes provides a non-trivial consistency check, although the routes play rather different practical roles. The $GOO{+}LOO$ cusp constraints can be imposed directly at the integrand level, whereas the $\mathrm{Tw2}{+}\mathrm{HS}$ route is tied more directly to the extracted CFT data and its analytic dependence on spin, but requires evaluating the conformal integrals to obtain finite-spin OPE data. Additionally, the $GG$- and $OO$-channel OPE provide overconstraints of the system, and thus furnish further independent checks.

\begin{table}[t]
\centering
\normalsize
\setlength{\tabcolsep}{-2pt}
\renewcommand{\arraystretch}{1.08}
\begin{tabular}{@{}p{0.72\columnwidth}r@{}}
\hline\hline
Constraint & Coeffs.\ fixed \\
\hline
$GOO$ cusp & 7 \\
$LOO$ cusp & 6 \\
$GG$-channel OPE & 1 \\
$OO$-channel OPE & 3 \\
$GO$-channel OPE & 2 \\
Integrated correlator ($\mathrm{Int}$) & 2 \\
Extraction of twist-two OPE data ($\mathrm{Tw2}$) & 7 \\
Harmonic-sum representability ($\mathrm{HS}$) & 4 \\
\hline
\noalign{\vskip 1pt}
\hline
Minimal sets & Coeffs.\ fixed \\
\hline
$GO+\mathrm{Int}+GOO+LOO$ & 13 \\
$GO+\mathrm{Int}+\mathrm{Tw2}+\mathrm{HS}$ & 13 \\
\hline\hline
\end{tabular}
\caption{Number of independent three-loop coefficients fixed by each individual constraint and by two minimal constraint sets, each leaving one parameter unfixed.}
\label{tab:3loopConstraintCount}
\end{table}

We denote the unfixed coefficient by $c$, and the correlator takes the following form 
\begin{align} \label{eq:T3c}
\!\!\!\!\! F_7(c) &= (2 c {+} \alpha ^2 {+} \alpha {-} 1 ) f^{(3)}_{1,4}(x_i) +cf^{(3)}_{1,1}(x_i) \, {+}  f^{(3)}_{1,2}(x_i)  \cr
& \,+ f^{(3)}_{1,3} (x_i) -2 (c {-} 2 \alpha {+} 2)\, [ f^{(3)}_{2,3}(x_i)-f^{(3)}_{3,3}(x_i) ] \cr 
& \,+ (c{+}\alpha ) \,  [ 2  f^{(3)}_{3,1} (x_i) -f^{(3)}_{2,1}(x_i) - f^{(3)}_{4,1}  (x_i)]  \, . 
\end{align}
We finally apply the 10-dimensional hidden symmetry to obtain the correlator $\langle G\,G\, \mathcal{O}_{k_1} \mathcal{O}_{k_2} \rangle$ at three loops; Unlike the simplest case \(k_1=k_2=2\), such generic four-point functions have nontrivial \(R\)-symmetry dependence, and their integrated correlators contain \(\zeta(3)^2\) in addition to \(\zeta(7)\), due to the non-planar \(f\)-graphs. As shown in Appendix~\ref{app:hidden}, requiring the \(\zeta(3)^2\) terms to vanish in each \(R\)-symmetry channel with even $k_1,k_2$ fixes \(c=(1-\alpha)/2\) \footnote{While this manuscript was under revision, Ref.~\cite{He:2026onf} appeared, presenting a computation of the large-spin scaling function for the maximal giant graviton ($\alpha=1$) with some assumptions of integrability. Their result coincides with the $\alpha=1$, $c=0$ specialization of~\eqref{eq:largeSpinUnfixed}, in agreement with the value $c=(1-\alpha)/2$ fixed by the hidden-symmetry constraint.}.
The three-loop correlator is then given by 
\begin{align} \label{eq:T3}
\!\!\!\!\! F_7 &= \alpha^2 f^{(3)}_{1,4}(x_i) + \frac{(1{-}\alpha)}{2} f^{(3)}_{1,1}(x_i) \, {+}  f^{(3)}_{1,2}(x_i)  \cr
& \,+ f^{(3)}_{1,3} (x_i) - 5 (1{-}\alpha) \, [ f^{(3)}_{2,3}(x_i)-f^{(3)}_{3,3}(x_i) ] \cr 
& \,+ \frac{(1{+}\alpha)}{2} \,  [ 2  f^{(3)}_{3,1} (x_i) -f^{(3)}_{2,1}(x_i) - f^{(3)}_{4,1}  (x_i)]  \, . 
\end{align}
We note that there are non-trivial contributions from the non-planar $f$-graphs, which are required not only to match the integrated correlator but also to ensure the correct large-spin behaviour of the OPE coefficients. 
When $\alpha=-1$, all the planar $f$-graphs have coefficient \(1\), so the planar sector acquires enhanced full permutation symmetry, as at two loops. 
Furthermore, the term that dominates in the limit $\alpha \to -\infty$ is $f_{1,4}^{(3)}(x_i)$, in agreement with the effective description of the very-heavy dual giant graviton limit~\cite{Brown:2025huy}.

\paragraph{Large-spin limit of OPE data.} 

Our result allows one to extract new OPE data. Here we focus on the OPE coefficient $d_S$ of two giant gravitons and a twist-two operator of spin \(S\). We defer the explicit result of $d_S$ at finite spin to Appendix~\ref{sec:OPEcoefficient} and concentrate on its large-\(S\) limit, which can be obtained from the harmonic-sum representation using the package \texttt{HarmonicSums}~\cite{Ablinger:2014rba}. The one-parameter family $F_7(c)$ at three loops, together with the one- and two-loop results, implies the following large-spin expansion:
\allowdisplaybreaks
\begin{align}
\label{eq:largeSpinUnfixed}
\ln& \big({d_S}/{d_{S,\mathrm{tree}}}\big)^2 =  
-a \big( \zeta(2){+} 2\ln 2\,\ln s\big) \nonumber \\[-2pt]
&\  {+} \frac{a^2}{20} \Big[
(23{-}7\alpha) \zeta(2)^2 {+} \big(20 \ln 2\,\zeta(2) {-} 30 \alpha \zeta(3)\big)\ln s \nonumber\\[-4pt]
&\quad +30\ln 2\,\zeta(3)
+10(1{-}\alpha) \zeta(2)(\ln s)^2
\Big] \nonumber \\[-4pt]
&\ {-}\frac{a^3}{840}\Big[
420\ln 2\,\zeta(2) \zeta(3) 
+2100 \ln 2\, \zeta(5) \nonumber \\
&\quad
+2\big(56c {-} 604 \alpha {+} 78\alpha^2 {+} 915 \big)\zeta(2)^3 \nonumber\\[1pt]
&\quad
+35 \big(18c {+} \alpha^2 {-} 88 \alpha {+} 57\big) \zeta(3)^2 \nonumber\\[1pt]
&\quad
+21\big(
44\ln 2\,\zeta(2)^2
+ 5(12c {-} \alpha^2 {-} 24\alpha {+} 21) \zeta(2)\zeta(3) \nonumber\\[1pt]
&\qquad\quad {-}50(3c {-} \alpha {-} \alpha^2 {+} 5) \zeta(5) \big) \ln s \nonumber \\[1pt]
&\quad
+168(1{-}\alpha)(9{-}\alpha) \zeta(2)^2(\ln s)^2 \nonumber \\[-1pt]
&\quad +70(1{-}\alpha)(15{-}\alpha) \zeta(3)(\ln s)^3  
\Big]
+ O(1/s) \, ,
\end{align}
where $\ln s := \ln S + \gamma_{\rm E}$, with $\gamma_{\rm E}$ Euler's constant. 
We note that remarkably, when $\alpha=1$ (i.e., for the full determinant operator), all the terms proportional to $(\log s)^2$ and $(\log s)^3$ vanish. This was first observed at two loops in~\cite{Jiang:2019xdz}. Our result shows that the cancellation of $(\log s)^k$ terms 
with $1 < k \leq \ell$ at $\ell$ loops occurs only for $\alpha=1$ 
and persists to three loops~\footnote{This large-spin behavior is consistent with 
an independent light-cone bootstrap analysis. We thank Frank Coronado 
for private communication on this point.}. At three loops, this cancellation relies crucially on the non-planar $f$-graphs in~\eqref{eq:T3}.

\section{Conclusion and outlook}

In this work we developed a bootstrap framework for mixed giant graviton correlators in \(\mathcal N=4\) super-Yang--Mills theory. Focusing on \(\langle GGOO\rangle\) in the leading large-\(N\) limit, we showed that a basis of labelled \(f\)-graphs, together with cusp and OPE constraints, localization data, and hidden-symmetry input, is sufficient to determine the correlator through three loops. For the sphere giant graviton correlator, our construction reproduces the known results through two loops and yields the three-loop correction, which passes nontrivial checks from integrated correlators and from the large-spin limit of the extracted OPE data.

Our analysis suggests several natural directions for future work. It would be important to place the hidden symmetry for generic giant graviton sectors on firmer ground~{\renewcommand{\thempfn}{cite_10Dsym_comment}\footnote{We note that \cite{Chen:2026yzm} proposed an alternative ten-dimensional hidden symmetry, which is beyond the consideration in this paper.}} and to extend the bootstrap to more general \(\langle GG O_{p_1} O_{p_2}\rangle\) correlators with nontrivial \(R\)-symmetry dependence. Another interesting direction is to move beyond the leading large-\(N\) limit, where finite-\(N\) effects should further illuminate the interplay between determinant operators, non-planar dynamics, and defect-like descriptions of giant gravitons. More broadly, the present results indicate that heavy BPS sectors admit a surprisingly rigid perturbative organization and may provide a useful bridge between localization, OPE bootstrap, and the emergent higher-dimensional structures of AdS/CFT.

It is also natural to proceed to higher-loop correlators. In this direction, the integrand-level nature of the cusp constraints should be especially useful, since extracting finite-spin OPE data from conformal integrals becomes substantially more demanding at higher loops. Another possible simplification comes from the observed correlation between the \(\alpha\)-dependence of an \(f\)-graph coefficient and the power of \(x_{12}^2\) in the corresponding labelled \(f\)-graph. In the examples studied here, if a graph carries a factor \((x_{12}^2)^n\), with \(n\geq -1\), then its coefficient is a polynomial in \(\alpha\) of degree at most \(n+1\). If this pattern persists at higher loops, it would allow one to start from a smaller ansatz,
\begin{equation}
c_{n}^{(\ell)}(\alpha)=\sum_{i=0}^{n+1}c_{n,i}^{(\ell)}\alpha^i\, ,
\end{equation}
where the \(c_{n,i}^{(\ell)}\) are rational numbers. For determining higher-loop correlators, the integrated-correlator results will again be important, and they are known to all orders~\cite{Brown:2025huy}. However, the periods of the relevant $f$-graphs have not been computed exhaustively beyond three loops~\footnote{Many of these $f$-graph periods at four loops, including non-planar ones, have been obtained in~\cite{Zhang:2024ypu} and were used to verify the localization computation for $\langle \mathcal{O}_2\mathcal{O}_2\mathcal{O}_2\mathcal{O}_2 \rangle$ beyond the planar limit.}. Higher-loop results would be especially interesting for clarifying some of the remarkable structures observed here, for instance, the large-spin behaviour at $\alpha=1$,  and the enhanced full permutation symmetry of the planar sector at $\alpha=-1$.

\section*{Acknowledgments}

 We would like to thank Augustus Brown, Frank Coronado, Alessandro Georgoudis, Zhongjie Huang, Xuhang Jiang, Yunfeng Jiang, Yu Wu, Shun-Qing Zhang, Yang Zhang, Xinan Zhou, for insightful discussions. We also would like to thank  Frank Coronado, Yunfeng Jiang, Yu Wu, and Yang Zhang for their comments on the draft. We are especially grateful to Xuhang Jiang for his help with the evaluation of the three-loop integrals.  S.H. is supported by the National Natural Science Foundation of China under Grant No. 12225510, 12447101, and by the New Cornerstone Science Foundation. C.S. is supported by the China Postdoctoral Science Foundation under Grant No. 2022TQ0346, and the National Natural Science Foundation of China under Grant No. 12347146. C.W. is supported by a Royal Society University Research Fellowship No. URF$\backslash$R$\backslash$221015 and a STFC Consolidated Grant, ST$\backslash$X00063X$\backslash$1 ``Amplitudes, strings \& duality".

\onecolumngrid
\setcounter{secnumdepth}{2}
\appendix

\section{$f$-graphs up to three loops}  \label{app:f-graphs}

\begin{figure}[t]
\centering
\begin{tabular}{@{}cccc@{}}
\includegraphics[width=0.22\textwidth]{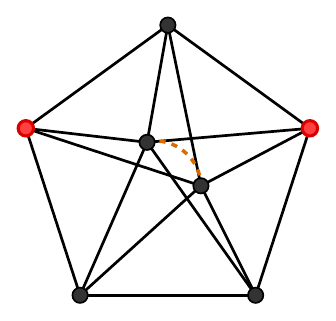} &\hspace{8pt}
\includegraphics[width=0.22\textwidth]{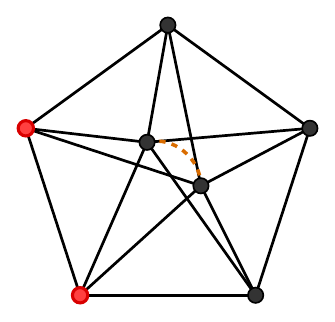} &\hspace{8pt}
\includegraphics[width=0.22\textwidth]{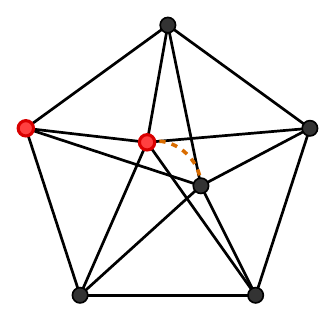} &\hspace{8pt}
\includegraphics[width=0.22\textwidth]{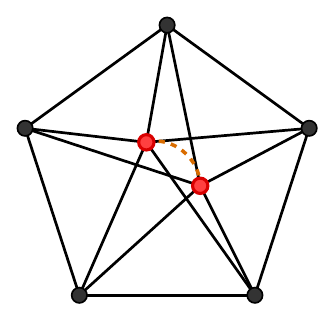} \\
$f_{1,1}^{(3)}$ &\hspace{8pt} $f_{1,2}^{(3)}$ &\hspace{8pt} $f_{1,3}^{(3)}$ &\hspace{8pt} $f_{1,4}^{(3)}$
\end{tabular}

\medskip
\begin{tabular}{@{}cccc@{}}
\includegraphics[width=0.22\textwidth]{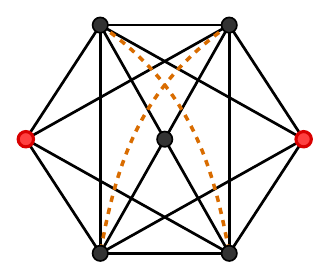} &\hspace{8pt}
\includegraphics[width=0.22\textwidth]{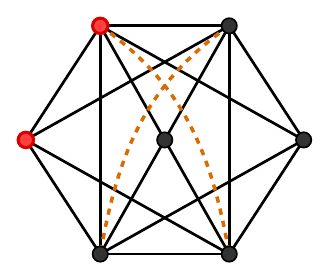} &\hspace{8pt}
\includegraphics[width=0.22\textwidth]{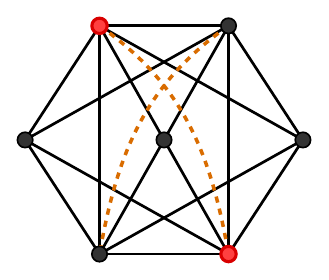} &\hspace{8pt}
\includegraphics[width=0.22\textwidth]{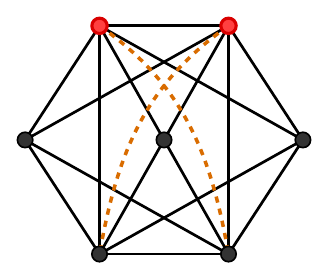} \\
$f_{2,1}^{(3)}$ &\hspace{8pt} $f_{2,2}^{(3)}$ &\hspace{8pt} $f_{2,3}^{(3)}$ &\hspace{8pt} $f_{2,4}^{(3)}$
\end{tabular}

\medskip
\begin{tabular}{@{}cccc@{}}
\includegraphics[width=0.22\textwidth]{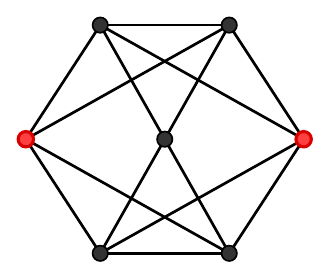} &\hspace{8pt}
\includegraphics[width=0.22\textwidth]{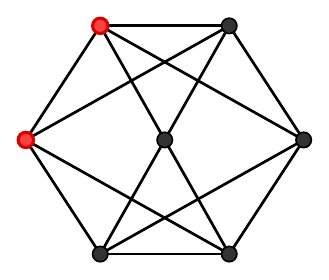} &\hspace{8pt}
\includegraphics[width=0.22\textwidth]{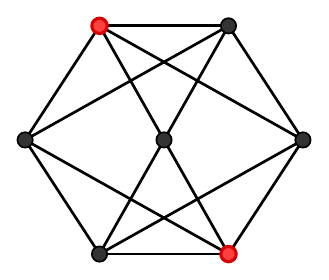} &\hspace{8pt}
\includegraphics[width=0.22\textwidth]{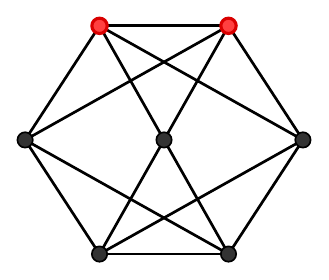} \\
$f_{3,1}^{(3)}$ &\hspace{8pt} $f_{3,2}^{(3)}$ &\hspace{8pt} $f_{3,3}^{(3)}$ &\hspace{8pt} $f_{3,4}^{(3)}$
\end{tabular}

\medskip
\makebox[\textwidth][c]{%
\begin{tabular}{@{}ccc@{}}
\includegraphics[width=0.22\textwidth]{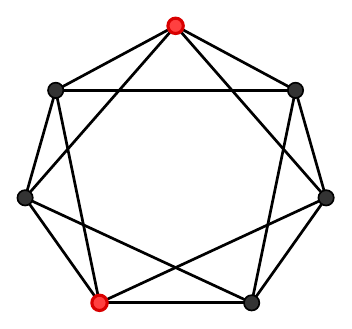} &\hspace{8pt}
\includegraphics[width=0.22\textwidth]{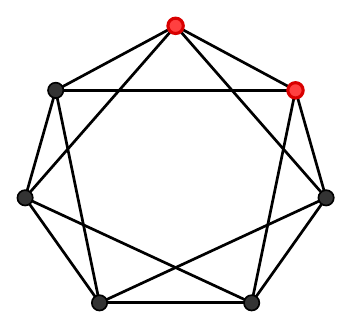} &\hspace{8pt}
\includegraphics[width=0.22\textwidth]{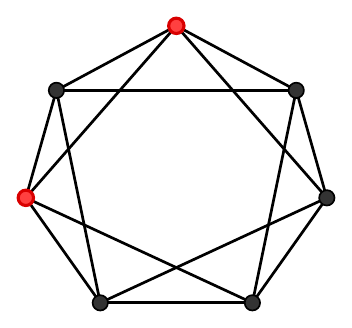} \\
$f_{4,1}^{(3)}$ &\hspace{8pt} $f_{4,2}^{(3)}$ &\hspace{8pt} $f_{4,3}^{(3)}$
\end{tabular}}
\caption{Three-loop labelled $f$-graphs organised by topology class. 
The graphs $f^{(3)}_{1,n}$ (first row) are planar; all others are 
non-planar. Highlighted red vertices mark the distinguished points $x_1$ 
and $x_2$ (the giant gravitons). Solid and dashed red lines denote 
propagators $1/x_{ij}^2$ and numerator factors $x_{ij}^2$, respectively.}
\label{fig:fgraph-3loop}
\end{figure}

Here we list all possible labelled $f$-graphs up to three loops; they form the integral basis for the giant graviton correlators studied in the main text. 
\paragraph{One loop.} At one loop, the only numerator $p^{(1)} (x_i)$ in the $f$-graph ansatz must be a degree-$0$ monomial, therefore, 
\begin{align} \label{eq:1-loop}
    f^{(1)}(x_i) = \frac{1}{\prod_{1\leq i < j \leq 5} x_{ij}^2}  \, .
\end{align}
\paragraph{Two loops.} At two loops, the only possible numerators $p^{(2)} (x_i)$  should be degree-$1$ monomial of $x_{ij}^2$, which leads to the following two possibilities
\begin{align} \label{eq:2-loop}
    p^{(2)}_1(x_i) &= {1\over 16} x_{12}^2 x_{34}^2 x_{56}^2    \, , \quad 
    p^{(2)}_2(x_i) = {1\over 4} x_{16}^2 x_{25}^2 x_{34}^2  \, ,
\end{align}
and
\begin{align}
    f^{(2)}_{n}(x_i)  = \frac{p^{(2)}_n(x_i) }{  \prod_{1 \leq i<j \leq 6} x^2_{i,j}} + P_{12;3456} \, . 
\end{align}
\paragraph{Three loops.}  There are 15 labelled $f$-graph integrals at three loops, which we denote as
\begin{align}
    f^{(3)}_{{\rm T},n}(x_i) = \frac{p^{(3)}_{{\rm T},n}(x_i)}{\prod_{1\leq i < j \leq 7} x_{ij}^2} +P_{12;34567} \, ,
\end{align}
where the numerators $p^{(3)}_{{\rm T},n}(x_i)$ are listed below according to the topologies. The ones that lead to planar graphs are 
\begin{align}
    p^{(3)}_{1,1}(x_i)=&\, \frac{1}{4} x_{1,2}^2 x_{1,5}^2 x_{2,3}^2 x_{3,4}^2 x_{4,5}^2 x_{6,7}^4\, , \qquad
    p^{(3)}_{1,2}(x_i)=\frac{1}{4} x_{1,3}^2 x_{1,5}^2
   x_{2,3}^2 x_{2,4}^2 x_{4,5}^2 x_{6,7}^4\, ,
\\
 p^{(3)}_{1,3}(x_i)=&\,\frac{1}{2} x_{1,5}^2 x_{1,6}^2 x_{2,7}^4 x_{3,4}^2
   x_{3,6}^2 x_{4,5}^2\, ,
\qquad
 p^{(3)}_{1,4}(x_i)= \frac{1}{20} x_{1,2}^4 x_{3,4}^2 x_{3,7}^2 x_{4,5}^2 x_{5,6}^2
   x_{6,7}^2 \, .
\end{align}
As we emphasized in the main text, even though we only consider the leading large-$N$ limit, the non-planar $f$-graphs do contribute to the giant graviton correlators. The numerators that give the non-planar graphs are: 
\begin{align}
  p^{(3)}_{2,1}(x_i)=&\, \frac{1}{16} x_{1,2}^2 x_{1,3}^2 x_{2,3}^2 x_{4,5}^4 x_{6,7}^4\,, \qquad   p^{(3)}_{2,2}(x_i)=\frac{1}{4} x_{1,3}^2 x_{1,4}^2 x_{2,5}^4
   x_{3,4}^2 x_{6,7}^4\,, 
   \\   p^{(3)}_{2,3}(x_i)=&\, \frac{1}{24}x_{1,2}^4 x_{3,4}^2 x_{3,5}^2 x_{4,5}^2 x_{6,7}^4\,, \qquad p^{(3)}_{2,4}(x_i)= \frac{1}{12}x_{1,5}^4
   x_{2,7}^4 x_{3,4}^2 x_{3,6}^2 x_{4,6}^2\, ; 
\end{align} 
\begin{align}
p^{(3)}_{3,1}(x_i)=&\,\frac{1}{16}x_{1,2}^2 x_{1,3}^2 x_{2,3}^2 x_{4,5}^2 x_{4,7}^2 x_{5,6}^2 x_{6,7}^2\, , \quad p^{(3)}_{3,2}(x_i)= \frac{1}{4} x_{1,3}^2
   x_{1,4}^2 x_{2,5}^2 x_{2,7}^2 x_{3,4}^2 x_{5,6}^2 x_{6,7}^2, \\ 
   p^{(3)}_{3,3}(x_i)=&\, \frac{1}{12} x_{1,2}^2 x_{1,7}^2
   x_{2,6}^2 x_{3,4}^2 x_{3,5}^2 x_{4,5}^2 x_{6,7}^2\,, \quad p^{(3)}_{3,4}(x_i)=  \frac{1}{24} x_{1,5}^2 x_{1,7}^2 x_{2,5}^2
   x_{2,7}^2 x_{3,4}^2 x_{3,6}^2 x_{4,6}^2 \, ; 
\end{align}
\begin{align}
  p^{(3)}_{4,1}(x_i)=&\, \frac{1}{2} x_{1,2}^2 x_{1,7}^2 x_{2,3}^2 x_{3,4}^2 x_{4,5}^2 x_{5,6}^2 x_{6,7}^2\, , \qquad  p^{(3)}_{4,2}(x_i)= \frac{1}{2} x_{1,3}^2
   x_{1,7}^2 x_{2,3}^2 x_{2,4}^2 x_{4,5}^2 x_{5,6}^2 x_{6,7}^2\, , \\ 
   p^{(3)}_{4,3}(x_i)=&\,  \frac{1}{2} x_{1,4}^2 x_{1,7}^2
   x_{2,3}^2 x_{2,5}^2 x_{3,4}^2 x_{5,6}^2 x_{6,7}^2 \, . 
\end{align}
The graphical representations are shown in Figure~\ref{fig:fgraph-3loop}. We note there is one Gram-determinant relation for the $f$-graph integrals at this order~\cite{Eden:2012tu}, implying one of the $f$-graphs is redundant. We used this fact in the main text in determining the three-loop correlators; more concretely, for convenience, we choose the basis where the coefficient of $f^{(3)}_{1,2}(x_i)$ is one. 

It was argued~\cite{Brown:2025huy} that in the very-heavy dual giant graviton limit, \(\alpha \to -\infty\), the giant graviton correlators admit an effective description related to moving \(\mathcal{N}=4\) SYM onto its Coulomb branch~\cite{Coronado:2025xwk}, analogous to the construction in~\cite{Caetano:2023zwe, Brown:2025cbz} in the so-called large-charge limit. In this limit, the correlator is expected to be dominated by the planar $f$-graphs with the highest powers of $x_{12}^2$ in the numerator, namely $f^{(2)}_1(x_i)$ at two loops and $f^{(3)}_{1,4}(x_i)$ at three loops. It is worth noting that these $f$-graphs can be expressed as sums of products of two ladder integrals~\cite{Brown:2024yvt}, which are the Feynman integrals dominating the large-charge limit; see, for example, \cite{Giombi:2020enj, Caetano:2023zwe, Brown:2025cbz}.
	
\section{Periods of $f$-graphs up to three loops} \label{app:periods}

In this section, we provide the periods for the $f$-graphs listed in Appendix~\ref{app:f-graphs}. For the first two loops, we have
\begin{align} 
\mathcal{P}(f^{(1)}) &= 6 \zeta(3) \, , \\
 \mathcal{P}(f^{(2)}_1) &=3 \times 20 \zeta(5) \, ,  \quad  \mathcal{P}(f^{(2)}_2) =12 \times 20 \zeta(5)  \, . 
\end{align}
The periods for the planar $f$-graphs at three loops are given by
\begin{align}
\mathcal{P}(f_{1,1}^{(3)}) =60\times 70 \zeta(7) \, , \quad \mathcal{P}(f_{1,2}^{(3)}) =60\times 70 \zeta(7) \, , \quad \mathcal{P}(f_{1,3}^{(3)}) =120\times 70 \zeta(7) \, ,\quad \mathcal{P}(f_{1,4}^{(3)}) = 12\times 70 \zeta(7) \, ,
\end{align}
These results were considered in \cite{Dorigoni:2021guq, Wen:2022oky} in the study of the integrated correlator of $\langle \mathcal{O}_2\mathcal{O}_2\mathcal{O}_2\mathcal{O}_2 \rangle$. 

For the non-planar $f$-graphs  at three loops, we have
\begin{align}
\mathcal{P}(f_{2,1}^{(3)}) &=15 \times (72 \zeta (3)^2-21 \zeta (7)) \, , \quad \mathcal{P}(f_{2,2}^{(3)}) =60\times  (72 \zeta (3)^2-21 \zeta (7)) \, , \\
 \mathcal{P}(f_{2,3}^{(3)}) &=10\times  (72 \zeta (3)^2-21 \zeta (7))  \, ,\quad \mathcal{P}(f_{2,4}^{(3)}) = 20 \times  (72 \zeta (3)^2-21 \zeta (7))  \, , \\
\mathcal{P}(f_{3,1}^{(3)}) &=15\times 36 \zeta (3)^2 \, , \quad \mathcal{P}(f_{3,2}^{(3)}) =60\times 36 \zeta (3)^2 \, , \quad \mathcal{P}(f_{3,3}^{(3)}) =20\times 36 \zeta (3)^2 \, ,\quad \mathcal{P}(f_{3,4}^{(3)}) = 10\times 36 \zeta (3)^2\, , \\
\mathcal{P}(f_{4,1}^{(3)}) &=\mathcal{P}(f_{4,2}^{(3)})  = \mathcal{P}(f_{4,3}^{(3)}) =120\times \frac{441}{8} \zeta(7) \, . 
\end{align}
The above results of the $f$-graph periods are obtained using  {\tt HyperlogProcedures}~\cite{HyperlogProcedures}. 

\section{Ten-dimensional hidden symmetry} \label{app:hidden}

We briefly review the proposal of~\cite{Chen:2025yxg, Wu:2025ott, Chen:2026ium}, also known as ten-dimensional hidden symmetry, according to which the full class of four-point functions $\langle G\,G\, \mathcal{O}_{k_1} \mathcal{O}_{k_2} \rangle$ can be packaged into a generating function in the leading large-$N$ limit, viewing the four-point functions as two-point correlators in the presence of a defect~\cite{cite_10Dsym_comment}.
In the weak-coupling expansion, the hidden-symmetry statement is that if one expresses the $\ell$-loop integrand in terms of $P_3 \cdot P_4$ and $P_i \cdot \mathbb{N} \cdot P_j$ for $i,j=3,4, \ldots, 4 + \ell$, with $P_k$ the six-dimensional embedding space vector, then the generating function is obtained by the following replacements:
\begin{align} \label{eq:prom}
P_3 \cdot P_4 &\to P_3 \cdot P_4 + Y_3 \cdot Y_4 \, , \\
P_i \cdot \mathbb{N} \cdot P_j &\to P_i \cdot \mathbb{N} \cdot P_j + Y_i \cdot \mathbb{M} \cdot Y_j\, ,
\end{align}
where the projectors are defined as
\begin{align}
\mathbb{N}_{AB} = \frac{P_{1,A}P_{2,B} + P_{2,A}P_{1,B}} {P_1 \cdot P_2}\, , \qquad \mathbb{M}_{IJ} = \delta_{IJ} -  \frac{Y_{1, I} Y_{2,J} + Y_{2, I} Y_{1,J}} {Y_1 \cdot Y_2} \, . 
\end{align}
This was studied at two loops in~\cite{Wu:2025ott}; here we are interested in the three-loop case. 
We note that the full $f$-graph ansatz can be recast in this defect form, with all dependence on $x_1, x_2$ absorbed into the projectors $\mathbb{N}$, making the defect picture manifest.
The resulting expression can then be uplifted to ten dimensions.
Such a reformulation, however, requires starting from an alternative ansatz formulated directly in the defect framework and relating its coefficients to those of the $f$-graph ansatz, which is beyond the scope of the current letter.
Here, it suffices to take advantage of the Gram identity to make a particular choice of integral basis, leading to the one-parameter three-loop expression presented in the main text.

For the purpose of fixing the unknown coefficient $c$ in the main-text expression, we are mainly interested in the cancellation of the $\zeta(3)^2$ terms appearing in the three-loop periods listed in Appendix~\ref{app:periods}. We find that for $\langle GG \mathcal{O}_{p} \mathcal{O}_{p+2k} \rangle$ with even $p$, the $\zeta(3)^2$  contribution to the integrated correlator in each \(R\)-symmetry channel is proportional to $(2c - 1 + \alpha)$. Requiring the cancellation of $\zeta(3)^2$ therefore fixes $c = (1 - \alpha)/2$, as stated in the main text.
For odd \(p\), after setting \(c=(1-\alpha)/2\), the \(\zeta(3)^2\) contribution has the common factor \((1-\alpha)[5(d_{13}d_{24}+d_{14}d_{23})-3d_{12}d_{34}]\). It is generically nonvanishing under \(d_{ij}=\gamma_i\gamma_j\), but vanishes in the maximal limit \(\alpha=1\).
This may not be surprising because, if the giant graviton insertions are viewed as defects, then $d_{12} d_{34}$ should naturally be treated differently from the other two structures. It would be interesting to understand how this pattern persists at higher loops.

\section{OPE data at three-loop order} 
\label{sec:OPEcoefficient}

In this appendix, we present three-loop OPE coefficients for the twist-two operators extracted from the three-loop correlator, and their large-spin behavior is studied in the main text.  In Table~\ref{tab:dc3FiniteSpin} we present the products of structure constants 
$d_S c_S$ for the first few even spins $S = 2, 4, 6$. The result is given for the one-parameter family $F_7(c)$ before fixing $c$ by hidden symmetry. The final answer can be obtained by setting $c = (1 - \alpha)/2$.
\begin{table}[th]
\centering
\newcommand{\dcformbox}[2]{\parbox[c][#1][c]{0.69\textwidth}{\centering #2}}
\begin{tabular}{c|c}
\hline
\parbox[c][0.32in][c]{0.07\textwidth}{\centering spin-$S$} &
\parbox[c][0.32in][c]{0.68\textwidth}{\centering $d_S c_S \big|_{\mathcal{O}(a^3)}$} \\
\hline
2 &
\dcformbox{0.4in}{\begin{align*}
-12+\frac{15\alpha-1}{8}\zeta(3)
-\frac{5(10-9\alpha+\alpha^2)}{8}\zeta(5)
\end{align*}} \\
\hline
4 &
\dcformbox{0.44in}{
\begin{align*}
&-\frac{899987671}{448084224}+\frac{c}{192}+\frac{43\alpha}{6912}
-\frac{14293-41145\alpha+882\alpha^2}{84672}\zeta(3) -\frac{25(10-9\alpha+\alpha^2)}{336}\zeta(5)
\end{align*}} \\
\hline
6 &
\dcformbox{0.75in}{
\begin{align*}
&\frac{-4863649611088+15984478125c+30417957075\alpha+988267500\alpha^2}{22769683200000}\\[2pt]
&-\frac{7(184046-447177\alpha+12375\alpha^2)}{52272000}\zeta(3)
-\frac{7(10-9\alpha+\alpha^2)}{1056}\zeta(5)
\end{align*}} \\
\hline
\end{tabular}
\caption{Finite-spin values of the three-loop coefficient $d_S c_S$ for $S=2,4,6$.}
\label{tab:dc3FiniteSpin}
\end{table}%

From the known results of $c_{S}$~\cite{Eden:2012rr}, we can determine the OPE coefficients $d_S$, which admits a closed form representation in terms of harmonic sums. 
In the coefficient functions below, we use the shorthand
\begin{align}
S_{a_1,\ldots,a_n}\equiv S_{a_1,\ldots,a_n}(S)\, .
\end{align}
The OPE coefficient up to three loops is given by
\allowdisplaybreaks
\begin{align} \label{eq:dS}
\left(\frac{d_S}{d_{S,\mathrm{tree}}}\right)^2
={}& \mathcal{P}_S \Big[1 + a\, d_{1,2} + a^2\, \big(d_{2,4} + d_{2,1} \zeta(3)\big)
 + a^3 \big(d_{3,6} + d_{3,3} \zeta(3) + d_{3,1} \zeta(5)\big)\Big]\, ,
\end{align}
where the prefactor is
\begin{align}
\mathcal{P}_S
={}& \frac{\Gamma(1 + 2 S)\Gamma(1 + S + \frac{\gamma_S}{2})^2}{\Gamma(1 + S)^2\Gamma(1 + 2 S + \gamma_S)}
+ \frac{\zeta(2)}{4} \Big(\gamma_S^2 + \big(S_1(S)-S_1(2S)\big) \gamma_S^3 \Big) - \frac{\gamma_S^3 \zeta(3)}{4}\, .
\end{align}
and the $d_{l,w}$'s are 
\begin{align}
d_{1,2}={}& -S_2\, , \\
d_{2,1}={}& -3\alpha S_1\, , \nonumber \\
d_{2,4}={}& \frac{1}{2} (1 - 2 \alpha) S_{-4}
{} + \frac{1}{2} (1 - \alpha) S_{-2}^{2}
{} + (3 + \alpha) S_{-3} S_{1}
{} + (1 + \alpha) S_{-2} S_{1}^{2}
{} - (1 + \alpha) S_{-2} S_{2}
{} + S_{2}^{2}
{} + (1 + \alpha) S_{1} S_{3} \nonumber\\
& {} + \frac{1}{2} (2 + \alpha) S_{4}
{} - (3 + \alpha) S_{-3,1}
{} - 2 (1 + \alpha) S_{1} S_{-2,1}
{} + (2 + \alpha) S_{2,-2}
{} + (1 - \alpha) S_{3,1}
{} + 2 (1 + \alpha) S_{-2,1,1}\, ,\\
d_{3,1}={}& -\frac{5}{2} (5 - 9 \alpha + \alpha^{2}) S_{1}\, ,\\
d_{3,3}={}& \frac{1}{2} (-17 + 28 \alpha - \alpha^{2}) S_{-3}
{} + \frac{1}{2} (-25 + 34 \alpha - \alpha^{2}) S_{-2} S_{1}
{} + \frac{1}{6} (-15 + 16 \alpha - \alpha^{2}) S_{1}^{3}
{} + 6 \alpha S_{1} S_{2}
\nonumber \\ &
{} + \frac{1}{6} (15 - 16 \alpha + \alpha^{2}) S_{3}
{} + (17 - 28 \alpha + \alpha^{2}) S_{-2,1}\, ,\\
d_{3,6}={}& \frac{1}{6} (-233 + 12 c + 444 \alpha + 15 \alpha^{2}) S_{-6}
{} + \frac{1}{4} (23 - 28 \alpha - \alpha^{2}) S_{-3}^{2}
{} + \frac{1}{2} (13 - 8 c - 24 \alpha + \alpha^{2}) S_{-4} S_{-2} 
{} - \frac{2}{3} (-2 + 3 \alpha) S_{-2}^{3}
\nonumber \\ & 
{} + \frac{1}{2} (59 + 4 c - 144 \alpha - 4 \alpha^{2}) S_{-5} S_{1}
{} + \frac{1}{2} (-43 + 40 \alpha + \alpha^{2}) S_{-4} S_{1}^{2}
{} + 6 (-1 + \alpha) S_{-2}^{2} S_{1}^{2}
{} + \frac{1}{6} (7 - 42 \alpha - \alpha^{2}) S_{-3} S_{1}^{3} 
\nonumber \\ & 
{} + \frac{1}{2} (53 - 118 \alpha - 4 \alpha^{2}) S_{-4} S_{2}
{} + \frac{1}{2} (26 - 8 c - 39 \alpha) S_{-2}^{2} S_{2}
{} + (-18 + 29 \alpha + \alpha^{2}) S_{-2} S_{1}^{2} S_{2}
{} + (18 - 29 \alpha - \alpha^{2}) S_{-2} S_{2}^{2}
\nonumber \\ & 
{} - S_{2}^{3} 
{} + (-17 + 2 c + 18 \alpha) S_{-2} S_{1} S_{-3} 
{} - 2 (6 + \alpha) S_{1} S_{2} S_{-3}
{} + \frac{1}{6} (-103 + 300 \alpha + 13 \alpha^{2}) S_{-3} S_{3} 
{} + \frac{1}{2} (-5 - \alpha) S_{2} S_{4} 
\nonumber \\ & 
{} + \frac{1}{6} (1 - 2 \alpha + \alpha^{2}) S_{1}^{3} S_{3}
{} + \frac{1}{12} (-11 + 4 \alpha - 5 \alpha^{2}) S_{3}^{2}
{} + \frac{3}{2} (-25 + 40 \alpha + \alpha^{2}) S_{-2} S_{4}
{} + \frac{1}{2} (1 - 8 \alpha + \alpha^{2}) S_{1}^{2} S_{4}
\nonumber \\ & 
{} + \frac{1}{2} (-30 + 20 \alpha + \alpha^{2}) S_{1} S_{5}
{} + \frac{1}{2} (-13 - 4 c + 8 \alpha) S_{6}
{} + (-53 - 4 c + 98 \alpha + 3 \alpha^{2}) S_{-5,1} 
\nonumber \\ & 
{} + \frac{1}{2} (9 + 4 c - 8 \alpha - \alpha^{2}) S_{-4,-2}
{} + (51 - 2 c - 60 \alpha - 2 \alpha^{2}) S_{1} S_{-4,1} 
{} + \frac{1}{2} (1 - 4 c + 40 \alpha + \alpha^{2}) S_{-4,2}
\nonumber \\ & 
{} - 2 (-7 + 2 c + 10 \alpha) S_{-2} S_{-3,1}
{} + (-11 + 34 \alpha + \alpha^{2}) S_{1}^{2} S_{-3,1}
{} + (-4 + 31 \alpha + \alpha^{2}) S_{2} S_{-3,1}
\nonumber \\ & 
{} + \frac{1}{2} (-19 + 96 \alpha + 3 \alpha^{2}) S_{1} S_{-3,2} 
{} + (-23 + 28 \alpha + \alpha^{2}) S_{-3} S_{-2,1}
{} + \frac{1}{3} (-19 + 30 \alpha + \alpha^{2}) S_{1}^{3} S_{-2,1}
\nonumber \\ & 
{} + \frac{1}{3} (-143 + 240 \alpha + 5 \alpha^{2}) S_{3} S_{-2,1}
{} + \frac{1}{2} (37 - 4 c - 48 \alpha - \alpha^{2}) S_{-2,1}^{2} 
{} + (55 - 88 \alpha - 3 \alpha^{2}) S_{-2,1}S_{1} S_{2} 
\nonumber \\ & 
{} -2 (2 + \alpha) S_{3} S_{1} S_{2} 
{} + \frac{1}{2} (71 - 120 \alpha - 3 \alpha^{2}) S_{3} S_{-2} S_{1} 
{} + \frac{1}{2} (3 - 4 c - 4 \alpha + \alpha^{2}) S_{-2,1} S_{-2} S_{1}
\nonumber \\ & 
{} + 2 (-7 + 2 c + 10 \alpha) S_{-2} S_{2,-2}
{} + (15 - 32 \alpha - \alpha^{2}) S_{1}^{2} S_{2,-2} 
{} + (-33 + 59 \alpha + 2 \alpha^{2}) S_{2} S_{2,-2}
\nonumber \\ & 
{} + (13 - 30 \alpha - \alpha^{2}) S_{3} S_{2,-1}
{} + (13 - 30 \alpha - \alpha^{2}) S_{-3} S_{2,1}
{} + (-13 + 30 \alpha + \alpha^{2}) S_{2,-1} S_{2,1}
{} + (-13 - 4 c - 8 \alpha) S_{3,-3} 
\nonumber \\ & 
{} + \frac{1}{2} (-103 - 4 c + 140 \alpha + 3 \alpha^{2}) S_{1} S_{3,-2}
{} + (-13 + 30 \alpha + \alpha^{2}) S_{2} S_{3,-1}
{} + (-1 + \alpha^{2}) S_{-2} S_{3,1}
{} - (1 {-} 2 \alpha {+} \alpha^{2}) S_{1}^{2} S_{3,1} 
\nonumber \\ & 
{} + (-1 + \alpha) S_{2} S_{3,1}
{} + \frac{1}{2} (-5 + 8 \alpha - 3 \alpha^{2}) S_{1} S_{3,2}
{} + \frac{1}{2} (61 - 4 c - 112 \alpha - 3 \alpha^{2}) S_{4,-2} 
{} - \frac{3}{2} (3 - 4 \alpha + \alpha^{2}) S_{1} S_{4,1}
\nonumber \\ & 
{} + (-1 + 2 \alpha - \alpha^{2}) S_{4,2} 
{} + 2 (-1 + \alpha) S_{5,1}
{} + (-59 + 4 c + 80 \alpha + 3 \alpha^{2}) S_{-4,1,1}
{} + (37 - 94 \alpha - 3 \alpha^{2}) S_{1} S_{-3,1,1} 
\nonumber \\ & 
{} - 3 (-11 + 30 \alpha + \alpha^{2}) S_{-3,1,2}
{} - 2 (-10 + 30 \alpha + \alpha^{2}) S_{-3,2,1} 
{} + \frac{1}{2} (49 - 4 c - 60 \alpha - \alpha^{2}) S_{1} S_{-2,1,-2}
\nonumber \\ & 
{} + 4 (-7 + 2 c + 10 \alpha) S_{-2} S_{-2,1,1} 
{} - 2 (-19 + 30 \alpha + \alpha^{2}) S_{1}^{2} S_{-2,1,1}
{} + 2 (-37 + 59 \alpha + 2 \alpha^{2}) S_{2} S_{-2,1,1} 
\nonumber \\ & 
{} + (-47 + 92 \alpha + 3 \alpha^{2}) S_{1} S_{2,-2,1}
{} + (13 - 30 \alpha - \alpha^{2}) S_{2} S_{2,-1,1}
{} + (13 - 30 \alpha - \alpha^{2}) S_{2} S_{2,1,-1}
\nonumber \\ & 
{} + (69 + 4 c - 96 \alpha - 3 \alpha^{2}) S_{3,-2,1}
{} + (13 - 30 \alpha - \alpha^{2}) S_{3,-1,2} 
{} + (9 + 4 c - 8 \alpha - \alpha^{2}) S_{3,1,-2}
\nonumber \\ & 
{} + 3 (1 - 2 \alpha + \alpha^{2}) S_{1} S_{3,1,1} 
{} + 2 (1 - 2 \alpha + \alpha^{2}) S_{3,1,2}
{} + 2 (1 - 2 \alpha + \alpha^{2}) S_{3,2,1}
{} + 2 (1 - 2 \alpha + \alpha^{2}) S_{4,1,1} 
\nonumber \\ & 
{} + 4 (-13 + 30 \alpha + \alpha^{2}) S_{-3,1,1,1}
{} + (9 + 4 c - 8 \alpha - \alpha^{2}) S_{-2,1,-2,1} 
{} + 6 (-19 + 30 \alpha + \alpha^{2}) S_{1} S_{-2,1,1,1}
\nonumber \\ & 
{} - 4 (-16 + 30 \alpha + \alpha^{2}) S_{2,-2,1,1}
{} + (-13 + 30 \alpha + \alpha^{2}) S_{2,-1,1,2} 
{} + (-13 + 30 \alpha + \alpha^{2}) S_{2,1,-1,2}
\nonumber \\ & 
{} - 4 (1 - 2 \alpha + \alpha^{2}) S_{3,1,1,1}
{} - 8 (-19 + 30 \alpha + \alpha^{2}) S_{-2,1,1,1,1}\, .
\end{align}

\twocolumngrid
\nocite{He:2026onf,Zhang:2024ypu,Chen:2026yzm}
\bibliography{ref-giant}

\end{document}